\documentclass[12pt]{cernart}
\usepackage{times}
\usepackage{epsfig}
\usepackage{amssymb}
\usepackage{cite}
\usepackage{colordvi}

\begin {document}

\begin{titlepage}
\docnum{CERN--PH--EP/2010--001}
\date{2 January 2010}

\vspace{1cm}

\title{\LARGE {\bf The  Spin-dependent Structure Function of the Proton \boldmath{$g_1^p$} 
         and a Test of the Bjorken Sum Rule
              }}
\vspace*{0.5cm}
\collaboration{COMPASS Collaboration}


\vspace{2cm}

\begin{abstract}

The inclusive   double-spin asymmetry, $A_1^p$, has been 
measured at COMPASS in deep-inelastic
polarised muon scattering off a large polarised NH$_3$ target.
The  data, collected  in the year 2007,
 cover the range $Q^2 > 1\,({\rm GeV}/c)^2$, $0.004<x<0.7$
and improve the statistical precision of $g_1^p(x)$ by a factor of two
in the region $x<0.02$. 
The new proton asymmetries are combined with 
those previously published for the deuteron
to extract the non-singlet spin-dependent structure function $g_1^{NS}(x,Q^2)$. 
The isovector  quark density, $\Delta q_3(x,Q^2)$, is evaluated from a
NLO QCD fit of  $g_1^{NS}$. 
The   first moment of $\Delta q_3$
is in good agreement with the value predicted by the Bjorken sum rule
and corresponds to a ratio of the axial and vector coupling constants 
$|g_A/g_V| = 1.28 \pm 0.07 ({\rm stat.}) \pm 0.10 ({\rm syst.})$. 
\\\\
Keywords: COMPASS; Deep inelastic scattering; Spin; Structure function; QCD analysis; Bjorken sum rule
\vfill
\submitted{(To be Submitted to Physics Letters B)}
\end{abstract}

\noindent
{{\large  COMPASS Collaboration}\\[\baselineskip]}
%
%
%
M.G.~Alekseev\Iref{turin_i},
V.Yu.~Alexakhin\Iref{dubna},
Yu.~Alexandrov\Iref{moscowlpi},
G.D.~Alexeev\Iref{dubna},
A.~Amoroso\Iref{turin_u},
A.~Austregesilo\IIref{cern}{munichtu},
B.~Bade{\l}ek\Iref{warsaw},
F.~Balestra\Iref{turin_u},
J.~Ball\Iref{saclay},
J.~Barth\Iref{bonnpi}
% ...
%
%

\hbox to 0pt {~}

%
%
\Instfoot{bielefeld}{Universit\"at Bielefeld, Fakult\"at f\"ur Physik, 33501 Bielefeld, Germany\Aref{f}}
\Instfoot{bochum}{Universit\"at Bochum, Institut f\"ur Experimentalphysik, 44780 Bochum, Germany\Aref{f}}
\Instfoot{bonniskp}{Universit\"at Bonn, Helmholtz-Institut f\"ur  Strahlen- und Kernphysik, 53115 Bonn, Germany\Aref{f}}
\Instfoot{bonnpi}{Universit\"at Bonn, Physikalisches Institut, 53115 Bonn, Germany\Aref{f}}
\Instfoot{brno}{Institute of Scientific Instruments, AS CR, 61264 Brno, Czech Republic\Aref{g}}
\Instfoot{calcutta}{Matrivani Institute of Experimental Research \& Education, Calcutta-700 030, India\Aref{h}}
\Instfoot{dubna}{Joint Institute for Nuclear Research, 141980 Dubna, Moscow region, Russia}
\Instfoot{erlangen}{Universit\"at Erlangen--N\"urnberg, Physikalisches Institut, 91054 Erlangen, Germany\Aref{f}}
\Instfoot{freiburg}{Universit\"at Freiburg, Physikalisches Institut, 79104 Freiburg, Germany\Aref{f}}
\Instfoot{cern}{CERN, 1211 Geneva 23, Switzerland}
\Instfoot{liberec}{Technical University in Liberec, 46117 Liberec, Czech Republic\Aref{g}}
\Instfoot{lisbon}{LIP, 1000-149 Lisbon, Portugal\Aref{i}}
\Instfoot{mainz}{Universit\"at Mainz, Institut f\"ur Kernphysik, 55099 Mainz, Germany\Aref{f}}
\Instfoot{miyazaki}{University of Miyazaki, Miyazaki 889-2192, Japan\Aref{j}}
\Instfoot{moscowlpi}{Lebedev Physical Institute, 119991 Moscow, Russia}
\Instfoot{munichlmu}{Ludwig-Maximilians-Universit\"at M\"unchen, Department f\"ur Physik, 80799 Munich, Germany\AAref{f}{k}}
\Instfoot{munichtu}{Technische Universit\"at M\"unchen, Physik Department, 85748 Garching, Germany\AAref{f}{k}}
\Instfoot{nagoya}{Nagoya University, 464 Nagoya, Japan\Aref{j}}
\Instfoot{praguecu}{Charles University, Faculty of Mathematics and Physics, 18000 Prague, Czech Republic\Aref{g}}
\Instfoot{praguectu}{Czech Technical University in Prague, 16636 Prague, Czech Republic\Aref{g}}
\Instfoot{protvino}{State Research Center of the Russian Federation, Institute for High Energy Physics, 142281 Protvino, Russia\Aref{l}}
\Instfoot{saclay}{CEA DAPNIA/SPhN Saclay, 91191 Gif-sur-Yvette, France}
\Instfoot{telaviv}{Tel Aviv University, School of Physics and Astronomy, 69978 Tel Aviv, Israel\Aref{m}}
\Instfoot{triest_i}{Trieste Section of INFN, 34127 Trieste, Italy}
\Instfoot{triest}{University of Trieste, Department of Physics and Trieste Section of INFN, 34127 Trieste, Italy}
\Instfoot{triestictp}{Abdus Salam ICTP and Trieste Section of INFN, 34127 Trieste, Italy}
\Instfoot{turin_u}{University of Turin, Department of Physics and Torino Section of INFN, 10125 Turin, Italy}
\Instfoot{turin_i}{Torino Section of INFN, 10125 Turin, Italy}
\Instfoot{turin_p}{University of Eastern Piedmont, 1500 Alessandria,  and Torino Section of INFN, 10125 Turin, Italy}
\Instfoot{warsaw}{So{\l}tan Institute for Nuclear Studies and University of Warsaw, 00-681 Warsaw, Poland\Aref{n} }
\Instfoot{warsawtu}{Warsaw University of Technology, Institute of Radioelectronics, 00-665 Warsaw, Poland\Aref{o} }
\Instfoot{yamagata}{Yamagata University, Yamagata, 992-8510 Japan\Aref{j} }
%
%
\Anotfoot{+}{Deceased}
\Anotfoot{a}{Also at IST, Universidade T\'ecnica de Lisboa, Lisbon, Portugal}
\Anotfoot{b}{On leave of absence from JINR Dubna}
\Anotfoot{c}{Also at Chubu University, Kasugai, Aichi, 487-8501 Japan$^{\rm j)}$}
\Anotfoot{d}{Also at KEK, 1-1 Oho, Tsukuba, Ibaraki, 305-0801 Japan}
\Anotfoot{e}{Also at GSI mbH, Planckstr.\ 1, D-64291 Darmstadt, Germany}
\Anotfoot{f}{Supported by the German Bundesministerium f\"ur Bildung und Forschung}
\Anotfoot{g}{Suppported by Czech Republic MEYS grants ME492 and LA242}
\Anotfoot{h}{Supported by SAIL (CSR), Govt.\ of India}
\Anotfoot{i}{Supported by the Portuguese FCT - Funda\c{c}\~ao para a Ci\^encia e Tecnologia grants POCTI/FNU/49501/2002 and POCTI/FNU/50192/2003}
\Anotfoot{j}{Supported by the MEXT and the JSPS under the Grants No.18002006, No.20540299 and No.18540281; Daiko Foundation and Yamada Foundation}
\Anotfoot{k}{Supported by the DFG cluster of excellence `Origin and Structure of the Universe' (www.universe-cluster.de)}
\Anotfoot{l}{Supported by Supported by CERN-RFBR grant 08-02-91009}
\Anotfoot{m}{Supported by the Israel Science Foundation, founded by the Israel Academy of Sciences and Humanities}
\Anotfoot{n}{Supported by Ministry of Science and Higher Education grant 41/N-CERN/2007/0}
\Anotfoot{o}{Supported by KBN grant nr 134/E-365/SPUB-M/CERN/P-03/DZ299/2000}
\end{titlepage}

%


In previous publications the COMPASS collaboration has presented
new accurate values of the longitudinal spin asymmetry of the deuteron,
$A_1^d$, covering a large range of $x$ ($0.004 < x < 0.7)$ in the
region of deep inelastic scattering (DIS)  $Q^2 > 1\,({\rm GeV}/c)^2$ \cite{a1d,g1d2006}.
These new values have led to an improved determination of the spin
structure function $g_1^d(x,Q^2)$ in the low $x$ region where only
the SMC measurements existed before \cite{smc}.
 The first moment of $g_1^d(x)$ 
has also provided a more accurate value for the matrix element of
the flavour singlet axial current $a_0 = 0.35 \pm 0.03 ({\rm stat}) \pm 0.05 ({\rm syst.})$
at $Q^2 = 3\,({\rm GeV}/c)^2$, confirming the rather small contribution 
of the quark spins to the nucleon spin. 

In this letter we present new COMPASS results for the 
inclusive double-spin asymmetry of
the proton, $A_1^p$  and for the
spin structure function $g_1^p(x,Q^2)$ measured in the same kinematic range. In combination with the
deuteron data, these results yield an evaluation
of the isovector quark density, $\Delta q_3(x,Q^2) = \Delta u - \Delta d$, and its first moment
which in turn provides a test of the Bjorken sum rule.

The proton data were collected in 2007 using a three target cell
configuration and the upgraded COMPASS spectrometer, as described in \cite{Sidis09}.
The polarised target material is the ammonia previously used in the
SMC experiment \cite{smc97}. The polarisation was about 90\% in absolute value,
 measured with a relative error of $\pm$ 2\%\cite{yuri}.
The  muon beam 
has  a natural polarisation of about $- 80$\%.
The energy of the incoming muons is constrained to be 
in the interval \mbox{$140<E_\mu<180$ GeV} and their polarisation is
known with a relative precision of $\pm$ 5\%.
All events used in the present analysis are required
to have a reconstructed primary interaction
vertex (defined by the incoming and the scattered muon)  
inside one of the target cells. In order to cancel out the muon flux 
normalisation in the asymmetry calculation, 
incident
muons are only accepted when their extrapolated trajectory 
crosses all three cells. 

For most events the trigger is based on a combination of hodoscope signals
defining the trajectory of the scattered muon. In addition to these "inclusive
triggers", low $x$ events are also selected by 
an additional condition
on the energy deposit in the hadron calorimeter, which is then used as a
"semi-inclusive trigger". At large $x$ and $Q^2$ most events are  selected
by  conditions on the calorimeter signal only, without 
any input from hodoscopes. For this "calorimeter-only trigger" as well as
for the semi-inclusive one, the presence of a reconstructed hadron trajectory
compatible with the calorimeter information is required.

The kinematic region is defined by requiring the photon virtuality 
$Q^2>1\,({\rm GeV}/c)^2$
and the fractional energy $y$ transfered
 from the beam muon to the virtual photon to be between 0.1 and 0.9.
The region which is  most affected by radiative corrections
is eliminated by the cut $y<0.9$.
The total sample  after all cuts amounts to 85.3 million events. 

The longitudinal virtual-photon proton asymmetry, $A_1^p$, is evaluated 
from  the numbers of events collected in the different target cells by
the method used in our previous analyses of deuteron data \cite{a1d,g1d2006}.
Neighbouring target cells are polarised in opposite directions and
data from both target spin orientations are thus recorded simultaneously.
The  lengths of the cells are chosen so that the two samples collected with
opposite spin orientations have in average the same acceptance, which limits
the risk of false asymmetries. The target spin directions are reversed once per
day by rotating the magnetic field and a few times per year by changing
the microwave frequencies used for dynamic nuclear polarisation. 
The asymmetries are calculated from the
numbers of events in cells with opposite spin orientations collected
before and after a field rotation so that flux and acceptance factors
cancel out. 

Radiative corrections are applied separately to the asymmetries obtained
for the inclusive triggers and to those obtained for the semi-inclusive
and calorimeter-only triggers because radiative 
elastic events contribute only to 
the former ones.
Another correction is applied to account for the polarisation of the $^{14}$N nucleus.
For this spin 1 object the correction is proportional to $A_1^d(x)$ and affected
by factors accounting for the number of $^{14}$N nuclei vs. H atoms, the alignment
of the proton spin vs. the  $^{14}$N spin and the ratio of  $^{14}$N to H
polarisations \cite{rondon}:
\begin{equation}
\Delta A_1^p(x) = \frac{1}{3} \cdot (-\frac{1}{3}) \cdot \frac{1}{6} \cdot 
\frac{\sigma_d(x)}{\sigma_p(x)} \cdot A_1^d(x).
\end{equation}
The corrections for the various  intervals of $x$ are given in the appendix.
They are of the order of 0.01 for $x > 0.35 $ and are mainly important for 
the evaluation of the first moment of the spin structure function 
$\Gamma_1^p(Q^2) = \int_0^1 g_1^p(x,Q^2) dx $. 

\begin{figure}[th] 
\centering 
\includegraphics[width=0.6\textwidth,angle=0,clip]{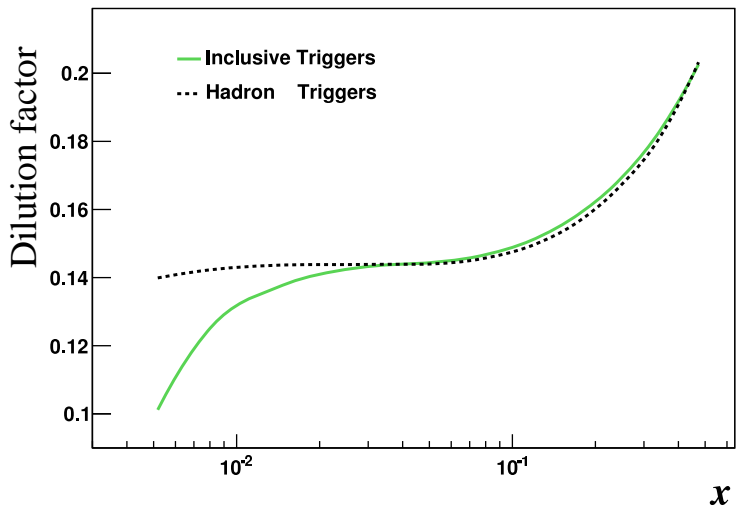} 
\caption{\small The dilution factor $f$ of the NH$_3$ polarised target as 
  a function of $x$ for inclusive and hadron triggers\cite{smc}. 
  The values of $f(x)$ are
  averaged over the $Q^2$  range of the corresponding triggers.}
\label{fig:dilfactor} 
\end{figure}

The target dilution factor is given by the ratio of the
cross-section for the polarisable protons to that of all nuclei  in a target cell. 
The values 
for the NH$_{3}$ target are shown 
in Fig.\,\ref{fig:dilfactor} as a function of $x$, for inclusive and 
hadron triggers. They are about 14\% in the medium $x$ region,
with a rise at large $x$ due to the reduced cross section on heavy targets
in this region, and a drop at low $x$ for inclusive triggers due to 
the contribution of radiative elastic events on the proton.

\begin{figure}[t] 
  \includegraphics[width=0.95\textwidth,clip]{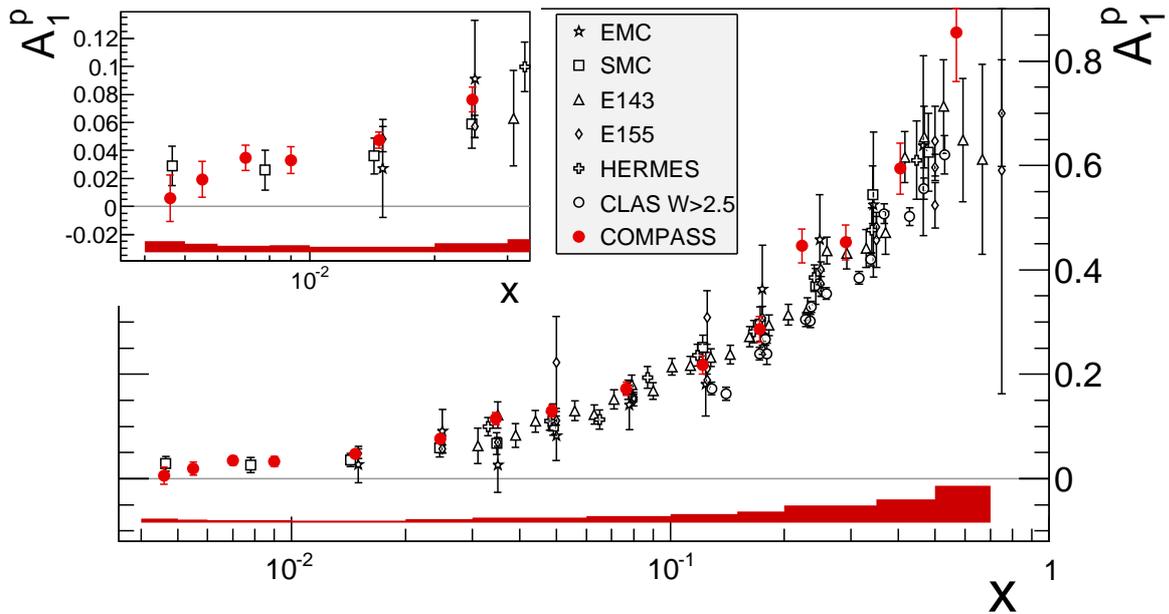} 
  \caption{\small The asymmetry $A_1^p(x)$ as measured by COMPASS and 
    previous results from EMC \cite{emc}, SMC \cite{smc}, HERMES \cite{semiinc4},
    SLAC E143 \cite{e143}, E155 \cite{e155} and CLAS \cite{CLAS} 
    at $Q^2>1\,({\rm GeV}/c)^2$.
    The cut $W>2.5$\,GeV has been applied to select DIS events in the CLAS data.
    Only statistical errors are shown with the data points. 
    The band at the bottom shows the estimated  
size of the  systematic errors for the COMPASS data.}
  \label{fig:asym} 
\end{figure}

\begin{figure}[t] 
  \includegraphics[width=0.95\textwidth,clip]{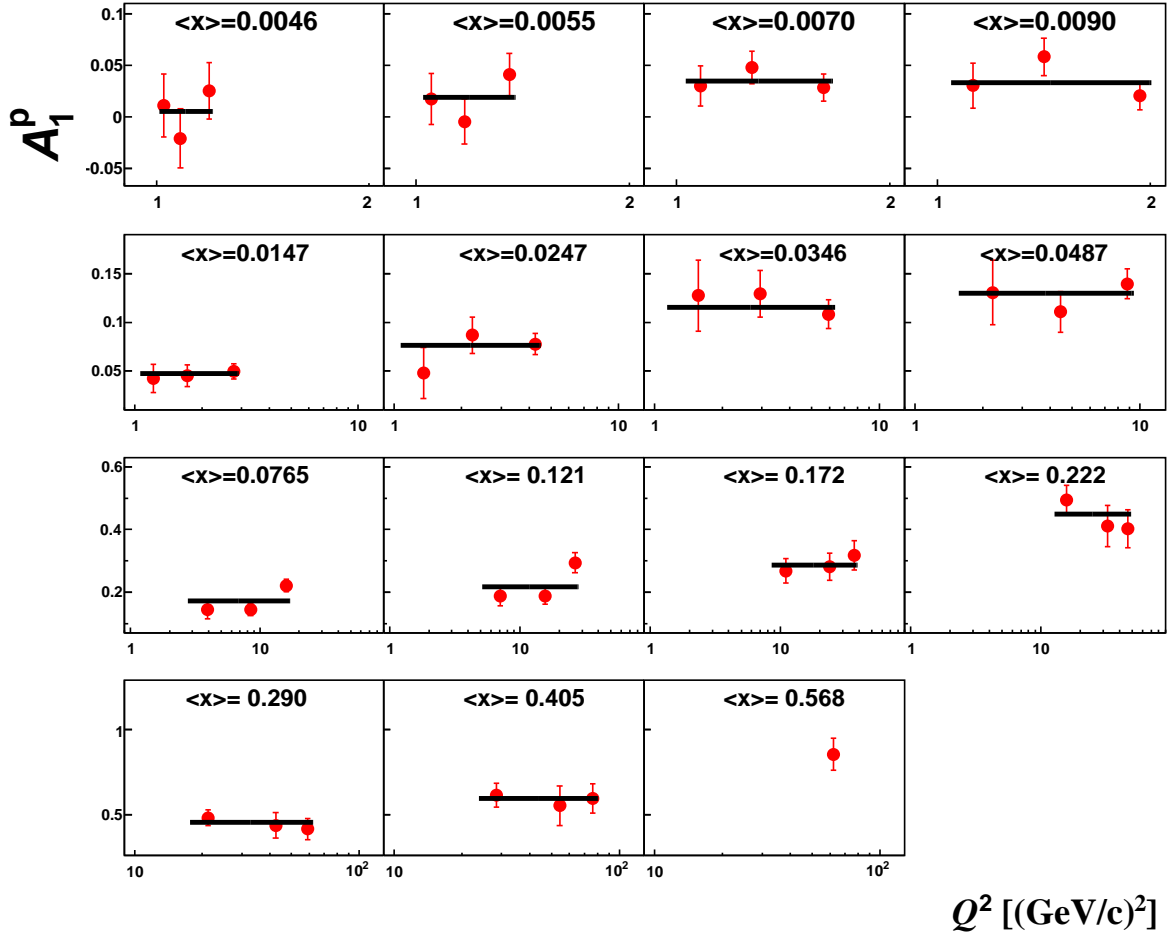} 
  \caption{\small Values of $A_1^p$ as a function of $Q^2$ in intervals 
     of $x$. The errors are statistical only. 
The solid lines show the results of fits to a constant.} 
  \label{fig:asym_vs_Q2} 
\end{figure}

The new values of $A_1^p(x)$ are shown in Fig.\,\ref{fig:asym} in
comparison with results from previous experiments. The $Q^2$ of different
points at any fixed value of $x$ varies considerably since the incident
energy of the various experiments ranges from 6 to 200 GeV. The fact 
that all results align reasonably well on a single curve illustrates
the well known observation that the $Q^2$ dependence of $A_1^p$ is very
weak in the DIS region. This is further illustrated in Fig.\,\ref{fig:asym_vs_Q2}
which shows $A_1^p(x,Q^2)$ as a function of $Q^2$ for the COMPASS data. 
No significant $Q^2$ dependence is observed in any interval of $x$.

The systematic errors of the COMPASS results for $A_1^p$ are 
shown by the band at the bottom of Fig.\,\ref{fig:asym} and listed in Table 1.
They contain the contributions
due to the uncertainties on the 
target polarisation, the beam polarisation, the dilution factor
and the ratio $R = \sigma_L/\sigma_T$ \cite{e143_R}  
used in the depolarisation factor, 
which are equal to 2, 5, 1 and at most 
3\%, respectively. Combined in quadrature, these
uncertainties amount to a systematic error of 
at most 6\% of the quoted value.
The error due to the neglect of the transverse asymmetry, $A_2^p$, is 
less than $0.002$ in the full range of $x$.
Another possible contribution to the systematic error is due to false
asymmetries generated by instabilities in some components of the 
spectrometer. Such effects have been searched for in faked configurations,
where the physics asymmetry does not contribute, and were found to be compatible
with zero. The comparison of results obtained with 
opposite orientations of the target field
also does not show any significant difference. The possible
error due to false asymmetries has been estimated by a statistical test
performed on the distribution of asymmetries extracted from 46 subsamples.
Time-dependent effects which would lead to a broadening of these distributions
were not observed. As a consequence 
the limit $\sigma_{syst} < 0.47 \sigma_{stat}$ was obtained at the level of one
standard deviation.
The different contributions to the systematic error are summarised in
Table \ref{tab:sys_error}.

\begin{figure}[here]
\epsfig{file=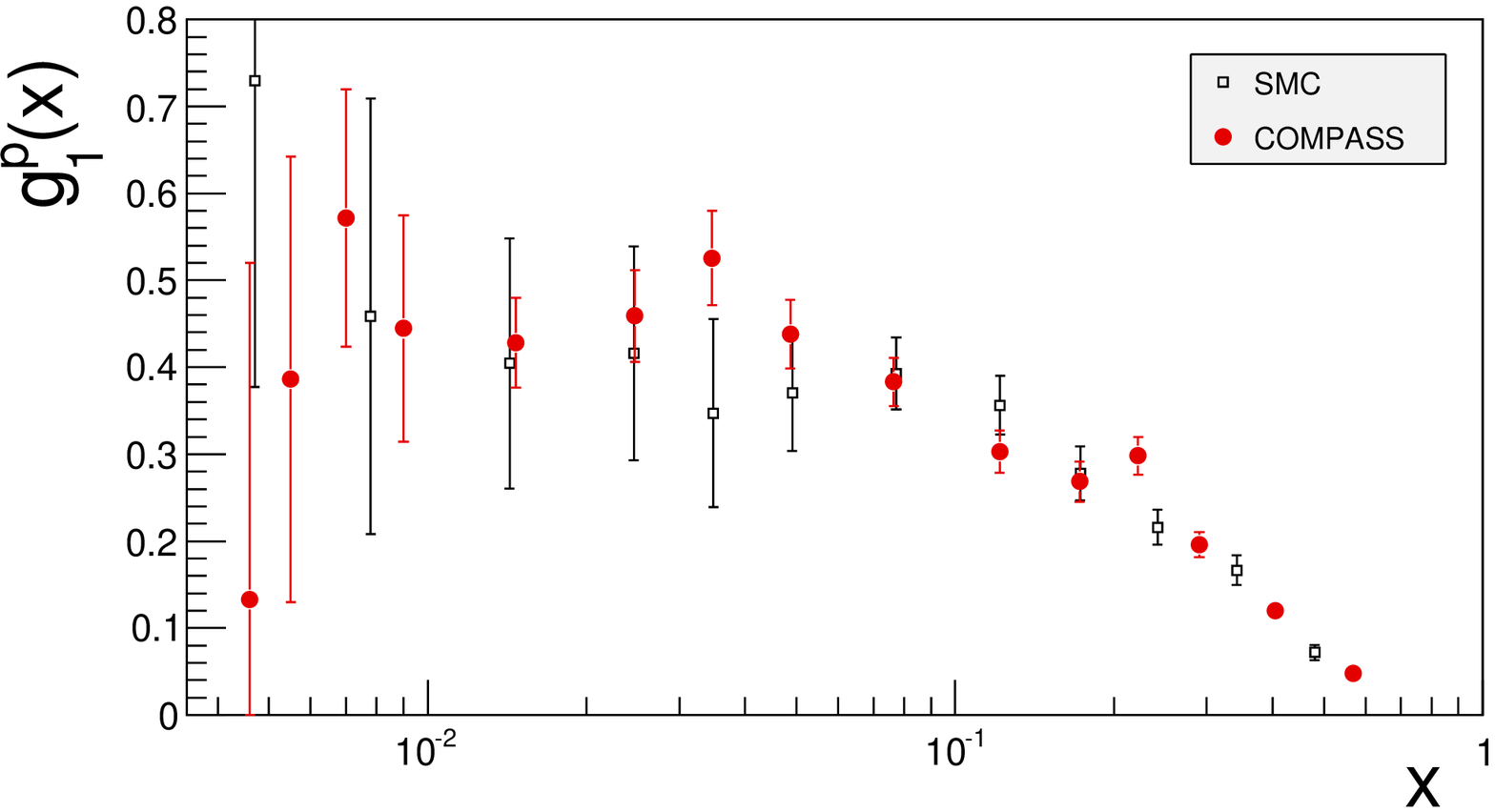,width=0.9\textwidth,clip=}
\caption{\small The spin structure function $g_1^p(x,Q^2)$ vs. $x$ 
 as measured by COMPASS at the
  $Q^2$ of each measured point. Previous results from SMC \cite{smc} are
shown for comparison. 
The errors are statistical only.}
\label{fig:g1p}
\end{figure}

The longitudinal spin structure function $g_1^p$ of the proton is obtained 
from $A_1^p$ by the relation
\begin{equation}
 g_1^p = \frac{F_2^p}{2 x (1+R)} A_1^p
\end{equation}
where $F_2^p$ is the spin independent structure function. The values obtained 
with the SMC parameterisation of the  world data on $F_2^p$ \cite{smc} 
and the parameterisation 
of $R$ already used in the depolarisation factor  are listed in
Table \ref{tab:a1_g1p} with their statistical and systematic errors.
They are also shown in Fig.\,\ref{fig:g1p} in comparison with the
SMC values. It can be seen that the COMPASS data improve the statistical
precision at least by a factor of two in the low $x$ region, covered only
by the two experiments shown here.  The
new data points 
are compatible with a constant $g_1^p(x)$ for 
$0.004 < x < 0.04$ and
do not show evidence  either for an increase or a decrease
when $x \rightarrow 0$.
 This observation remains valid when the data points
are moved to a common $Q^2$ 
according to the fits quoted in Ref.\cite{g1d2006} and the constant value
is found to be 
$g_1^p = 0.48 \pm 0.03 ({\rm stat.}) \pm 0.04 ({\rm syst.})$ at $Q^2 = 3 ({\rm GeV}/c)^2$.

\begin{table}[here]
\centering
\begin{tabular}{|c|c|c|c|c|}
\hline  \hline
$x$ range & $\langle x \rangle$ & $\!\!\langle Q^2 \rangle$ [$({\rm GeV}/c)^2$]$\!\!$  & $A_1^p$ & $g_1^p$ \\
\hline
0.004 - 0.005 & 0.0046 &  1.10 & $0.006 \pm 0.017 \pm 0.008$ & $0.133 \pm 0.389 \pm 0.183$\\
0.005 - 0.006 & 0.0055 &  1.20 & $0.019 \pm 0.013 \pm 0.006$ & $0.385 \pm 0.256 \pm 0.123$\\
0.006 - 0.008 & 0.0070 &  1.37 & $0.035 \pm 0.009 \pm 0.005$ & $0.571 \pm 0.147 \pm 0.079$\\
0.008 - 0.010 & 0.0090 &  1.59 & $0.033 \pm 0.010 \pm 0.005$ & $0.445 \pm 0.130 \pm 0.067$\\
0.010 - 0.020 & 0.0147 &  2.14 & $0.047 \pm 0.006 \pm 0.004$ & $0.427 \pm 0.052 \pm 0.036$\\
0.020 - 0.030 & 0.0247 &  3.24 & $0.076 \pm 0.009 \pm 0.006$ & $0.459 \pm 0.053 \pm 0.037$\\
0.030 - 0.040 & 0.0346 &  4.36 & $0.115 \pm 0.012 \pm 0.009$ & $0.525 \pm 0.054 \pm 0.041$\\
0.040 - 0.060 & 0.0487 &  6.05 & $0.130 \pm 0.012 \pm 0.010$ & $0.438 \pm 0.039 \pm 0.032$\\
0.060 - 0.100 & 0.0765 &  9.42 & $0.172 \pm 0.013 \pm 0.012$ & $0.383 \pm 0.028 \pm 0.026$\\
0.100 - 0.150 & 0.122 & 14.9 & $0.218 \pm 0.017 \pm 0.015$ & $0.303 \pm 0.024 \pm 0.021$\\
0.150 - 0.200 & 0.172 & 20.9 & $0.286 \pm 0.024 \pm 0.020$ & $0.268 \pm 0.023 \pm 0.019$\\
0.200 - 0.250 & 0.222 & 26.7 & $0.446 \pm 0.032 \pm 0.030$ & $0.298 \pm 0.022 \pm 0.020$\\
0.250 - 0.350 & 0.290 & 34.6 & $0.453 \pm 0.033 \pm 0.032$ & $0.196 \pm 0.014 \pm 0.013$\\
0.350 - 0.500 & 0.405 & 47.1 & $0.594 \pm 0.049 \pm 0.043$ & $0.120 \pm 0.010 \pm 0.008$\\
0.500 - 0.700 & 0.568 & 62.1 & $0.855 \pm 0.094 \pm 0.068$ & $0.048 \pm 0.005 \pm 0.004$\\
\hline  \hline
\end{tabular}
\caption{\small
  Values of $A_1^p$ and $g_1^p$ 
  as a function of $x$ with the corresponding average value of $Q^2$.
  The first error is statistical, the second one systematical.}
\label{tab:a1_g1p}
\end{table}

\begin{table}[here]
\begin{center}

\begin{tabular}{|l|c|c|}
\hline
               Beam polarisation     & $\Delta P_b/P_b$ & $0.04/0.8=5.0\%$ \\
 Target polarisation   & $\Delta P_t/P_t$ & 2\% \\
 Depolarisation factor & $\Delta D(R)/D(R)$ & 2.0 -- 3.0 \% \\
 Dilution factor & $\Delta f/f$    & 1 \% \\
Total             &    $\Delta A_1^{mult}$ & $\simeq 0.06 A_1$ \\
\hline
\hline
 Transverse asymmetry  & ${\eta}\cdot A_2$ & $< 2.0 \times 10^{-3}$ \\
 Rad. corrections& $\Delta A_1^{RC}$ &  $10^{-4}-10^{-3}$\\
 False asymmetry   & $A_{false}$ & $<0.47\cdot \Delta A_1^{stat}$ \\
\hline
\end{tabular}
\end{center}
\caption{\small Decomposition of the systematic error of $A_1^p$ into multiplicative (top)
   and additive (bottom) contributions.}
\label{tab:sys_error}
\end{table}

The non-singlet spin structure function
\begin{equation}
g_1^{NS}(x,Q^2) = g_1^p(x,Q^2) - g_1^n(x,Q^2)
\end{equation}
is of special interest because its $Q^2$ dependence is decoupled  from
the singlet and the gluon spin densities:
\begin{equation}
g_1^{NS}(x,Q^2) = \frac{1}{6} \int_x^1 \frac{dx'}{x'} C^{NS} \Bigl ( \frac{x}{x'},
\alpha_s(Q^2) \Bigr ) \Delta q_3(x',Q^2) 
\end{equation}
where $C^{NS}$ is a Wilson coefficient function and $\Delta q_3$ the
isovector spin density.
Consequently a fit of the $Q^2$ evolution of $g_1^{NS}$ requires only a small 
number of parameters to describe the shape of 
$\Delta q_3(x)$ at some reference $Q^2$.
According to the Bjorken sum rule
the integral of $g_1^{NS}$ at any fixed $Q^2$ is proportional
to the ratio $g_A/g_V$ of the axial and vector coupling constants and given by 
the relation
\begin{equation}
  \Gamma_1^{NS}(Q^2) = \frac{1}{6} \Bigl| \frac{g_A}{g_V} \Bigr| C_1^{NS}(Q^2)
  \label{Eq:BSR}
\end{equation}
where  the  non-singlet coefficient function  $C_1^{NS}(Q^2)$ 
has been calculated in perturbative QCD up to the third order in 
$\alpha_s(Q^2) $ \cite{Larin}.
The comparison of the  value of  $|g_A/g_V|$ obtained from the data with the one
derived from neutron $\beta$ decay ($|g_A/g_V| = 1.2694\pm0.0028$\cite{amsler}) 
thus  provides a test  of the Bjorken sum rule, free of systematic errors arising
from uncertainties on the gluon helicity distribution. 

In the present analysis, the values  of $g_1^{NS}$ are obtained as
\begin{equation}
g_1^{NS} (x,Q^2) = 2 \Bigl[ g_1^p(x,Q^2) - \frac{g_1^d(x,Q^2)}{1 - 1.5 \omega_D} \Bigr]
\end{equation}
where the values of $g_1^d$ are taken from Ref.\cite{g1d2006} and the
deuteron D-state probability, $\omega_D$,  is $ 0.05\pm 0.01$ \cite{machleidt}.

The proton and deuteron data have been obtained at the same $(x,Q^2)$ points,
which avoids the need of interpolation.
A correction has however to be applied to the deuteron data for the admixtures
of $^7$Li  and $^1$H in the $^6$LiD target material, which was not taken
into account in Ref.\cite{g1d2006}.
The ratios of isotopes $^7$Li/$^6$Li and H/D were found to be 4.4 \% and 0.5 \% 
respectively \cite{neliba} and $^7$Li and $^1$H  are both polarised at more than 90\% \cite{ball}.
The resulting corrections to $A_1^d$ are given in the appendix.
They are negligible at low $x$ but reach $0.015$ 
at $x = 0.50$ and reduce the first moment $\Gamma_1^d$ by 0.002.

\begin{figure}[tb]
\epsfig{file=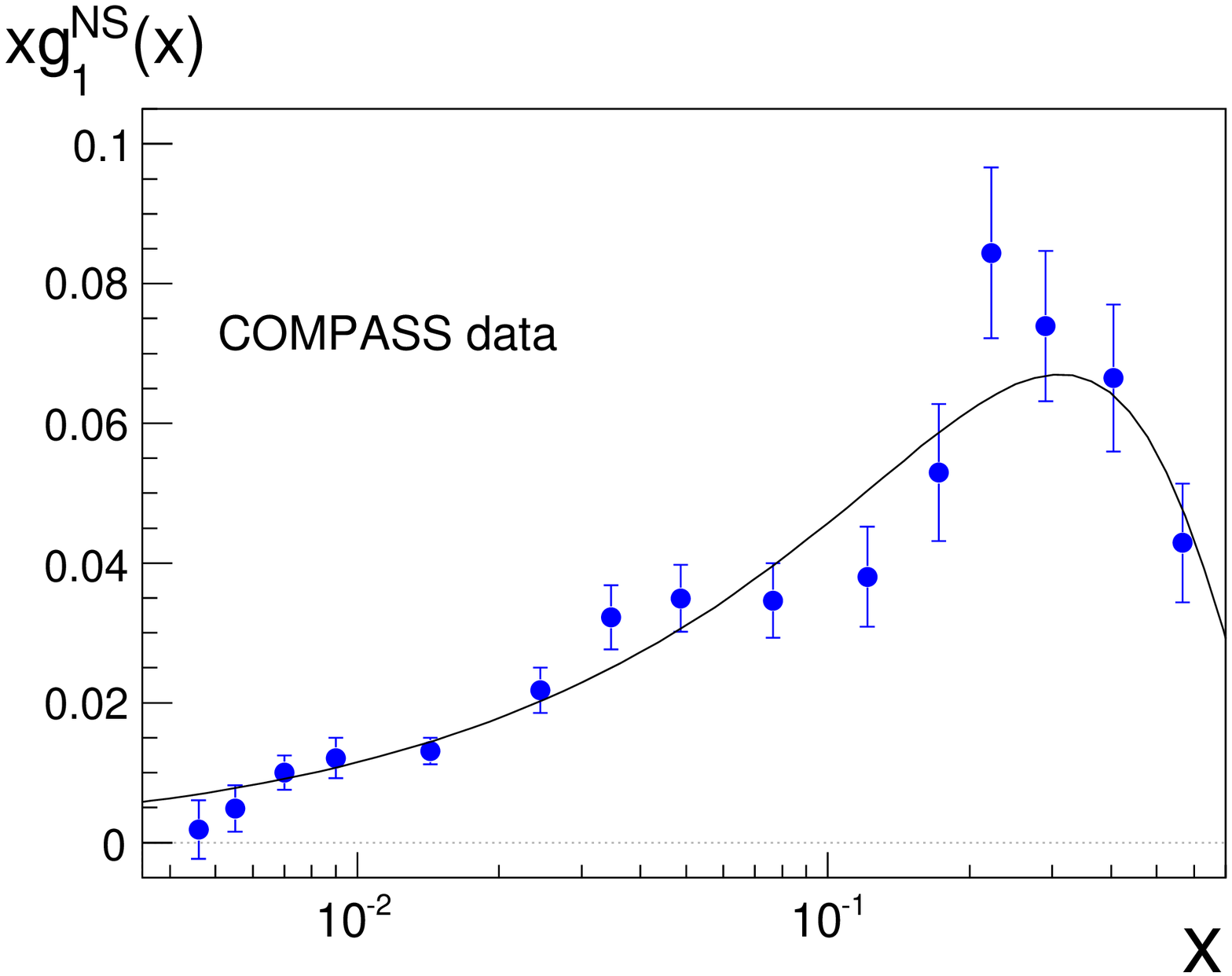,width=0.49\textwidth,clip=}
\hfill
\epsfig{file=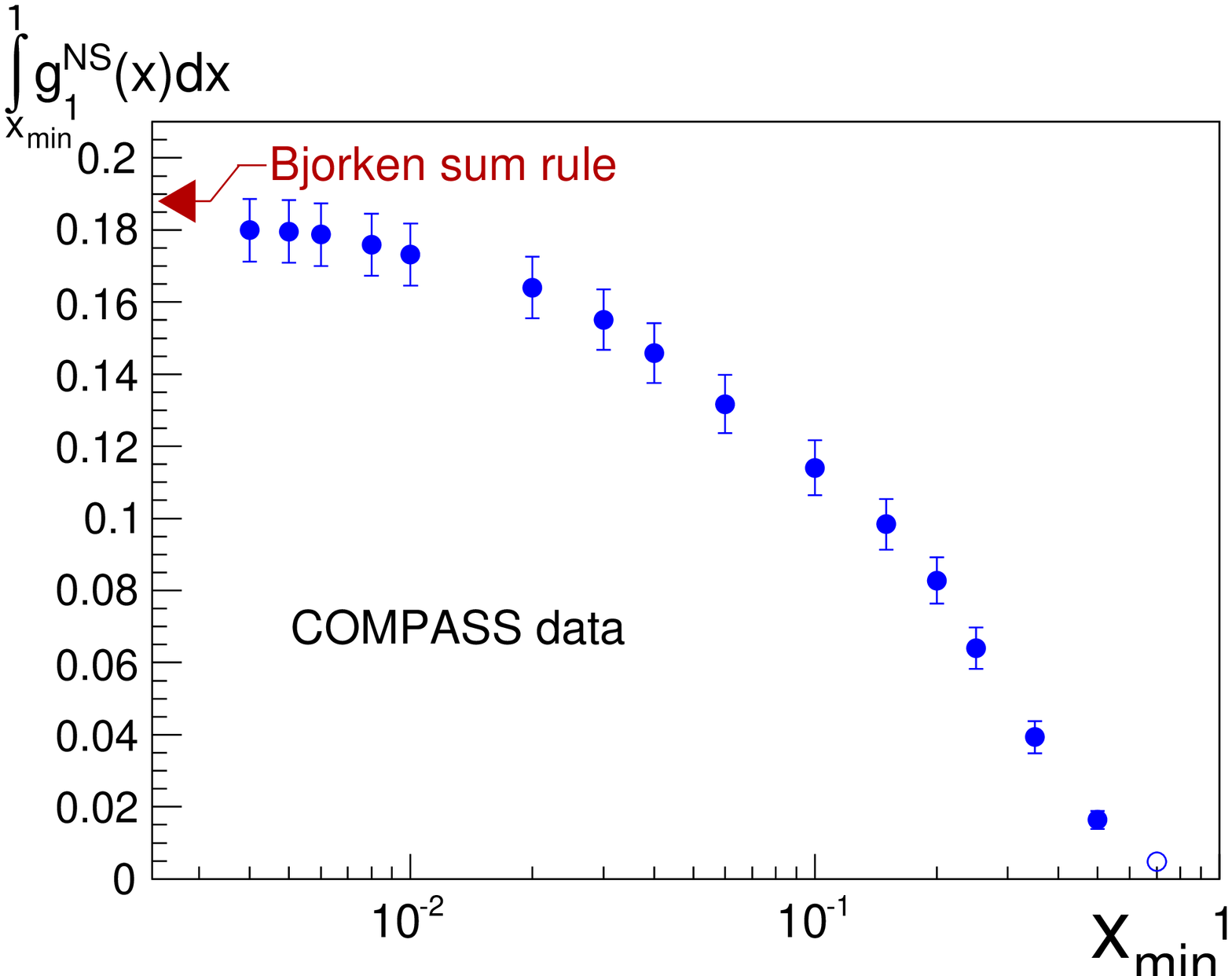,width=0.49\textwidth,clip=}
\caption{\small
  Left:  Values of $g_1^{NS}(x)$ at $Q^2=3\,({\rm GeV}/c)^2$,
  derived from the COMPASS measurements of $A_1^p$ and $A_1^d$  and 
  result of a three  parameter QCD fit at NLO. 
  The errors are statistical only.
  Right: $\int_{x_{min}}^1 g_1^{NS} dx$ as a function of $x_{min}$ as
  obtained from the COMPASS data points. 
  The open circle at $x=0.7$ is obtained from the fit.
  The arrow on the left side shows the value expected for the full range 
  $0 < x < 1$ with $|g_A/g_V| = 1.269$ \cite{amsler}.}
\label{fig:g1NS}
\end{figure}

In the present analysis, $Q^2 = 3\,({\rm GeV}/c)^2$ has been taken as
reference $Q^2$ and the following parameterisation has been used for $\Delta q_3$:
\begin{equation}
\Delta q_3(x) = \eta_3 
\frac{x^{\alpha_3} (1 - x)^{\beta_3} }{\int_0^1 x^{\alpha_3} (1 - x)^{\beta_3} dx } \,.
\end{equation}
As for our previous analysis \cite{g1d2006} the QCD fit at NLO
of the COMPASS values of $g_1^{NS}$ to Eq.(4) 
has been performed with two 
different programs, the first one working in the ($x$,$Q^2$) space \cite{smc_qcd},
the second one in the space of moments \cite{Sissakian}. Both programs give 
the same values of the fitted parameters and similar $\chi^2$-probabilities.
The fitted parameters obtained  are listed in Table \ref{tab:NS_fit}.
The exponent $\alpha_3$ is in the range of Regge pole predictions\cite{bass}
and the integral $\eta_3$ is in excellent agreement with the Bjorken
sum rule prediction $|g_A/g_V|$.
The fitted distribution of $x g_1^{NS}(x)$ and the data points
moved to the reference $Q^2$ are shown in Fig.\,\ref{fig:g1NS} (left). 
\begin{table}
\begin{minipage}[t]{0.4\textwidth}
\centering
\begin{tabular}{|c|c|}
\hline
Param. & Value \\
\hline
$\eta_3$  & $ 1.28 \pm 0.07$ \\
$\alpha_3$& $-0.22 \pm 0.07$ \\
$\beta_3$ & $ 2.2  \pm ^{0.5}_{0.4} $ \\
\hline
$\chi^2/$ND & 14.4/12  \\
Prob. &   0.27 \\
\hline
\end{tabular}
\caption{\small
  Results of the fits of $\Delta q_3(x)$ at \mbox{$Q^2= 3\,({\rm GeV}/c)^2$}.}
\label{tab:NS_fit}
\end{minipage}
\hfill
\begin{minipage}[t]{0.58\textwidth}
\centering
\begin{tabular}{|rcl|l|}
\hline
\multicolumn{3}{|c}{$x$ range}  & \multicolumn{1}{|c|}{$\Gamma_1^{NS}$} \\
\hline
    0 &$\!\!\!-\!\!\!$& 0.004 &  $0.0098$          \\
0.004 &$\!\!\!-\!\!\!$& 0.7   &  $0.175\pm0.009\pm0.015$ \\ 
  0.7 &$\!\!\!-\!\!\!$& 1.0   &  $0.0048$          \\
\hline
    0 &$\!\!\!-\!\!\!$& 1     &  $0.190\pm0.009\pm0.015$ \\
\hline
\end{tabular}
\caption{\small
  First moment $\Gamma_1^{NS}$ at $Q^2 =3\,({\rm GeV}/c)^2$\, from 
  the COMPASS data points. The contributions from  the unmeasured regions 
  are estimated from the NLO fit to $g_1^{NS}$; their errors are negligible.}
\label{tab:NS_integ}
\end{minipage}
\end{table}

The integral of $g_1^{NS}$ has also been evaluated  from the
measured values in the range $0.004 < x < 0.7$ with additional
low and high $x$ contributions taken from the fit (Table \ref{tab:NS_integ}).
It is observed that about
92\% of the first moment $\Gamma_1^{NS}$ comes from the measured region.
The dependence of the first moment of $g_1^{NS}$ on its lower limit is shown
in Fig.\,\ref{fig:g1NS} (right). As already observed in the HERMES analysis
\cite{semiinc4}, the integral does not saturate at $x \approx 0.01-0.02$
while the value obtained at the lowest $x$ accessible in the present
analysis ($0.180 \pm 0.009$) is less than one standard deviation below 
the value expected from the Bjorken sum rule ($0.188$).
The value of $|g_A/g_V|$ 
derived from the value of $\Gamma_1^{NS}$ by Eq.(\ref{Eq:BSR})
is identical to the one obtained from the fit and confirms the validity 
of the Bjorken sum rule with a statistical precision of 5\%.

The dominant systematic error  on this result is due to the uncertainty of
5\% on the beam polarisation, which is common to the proton and deuteron
data and therefore translates directly into a 5\% error
on $|g_A/g_V|$. Other contributions due to the target polarisation and the 
dilution factor are estimated to be
$\pm 0.04$  and $\pm 0.06$ for the proton  and deuteron terms, respectively. 
The resulting  systematic error is  $\pm 0.10$. The errors related to the fit
or to the evolution of the data to a common $Q^2$ are found to be negligible.
In particular, it was checked that the same value of $g_A/g_V$ is obtained 
when the reference $Q^2$ is 1.0, 3.0 or 10.0 $({\rm GeV}/c)^2$ although the 
exponent $\alpha_3$ varies from $-0.15$ to $-0.28$ when $Q^2$ is moved
from 1 to 10 and the shape of $g_1^{NS}(x)$ thus becomes quite different. 

The test of the Bjorken sum rule performed in the present analysis 
of the COMPASS proton and deuteron data is thus
mainly limited by systematics: 
\begin{equation}
 |g_A/g_V| = 1.28 \pm 0.07 ({\rm stat.}) \pm 0.10 ({\rm syst.}),
\end{equation}
to be compared with the value $1.2694 \pm 0.0028$ derived from neutron
$\beta$ decay\cite{amsler}.\\ 
The COMPASS value of $\Gamma_1^{NS}$is in good agreement with the one obtained
by the SMC ($0.198 \pm 0.023$ at $Q^2$ = 10 $({\rm GeV}/c)^2$)\cite{smc} and improves
the statistical precision by a factor 2.5. The cumulative integral of $g_1^{NS}$
truncated at $x_{min} = 0.021$ is equal to
$0.1583 \pm 0.0085 ({\rm stat.}) \pm 0.014 ({\rm syst.})$,
in good agreement with the HERMES value  
of $0.1484 \pm 0.0055 ({\rm stat.}) \pm 0.016 ({\rm syst.})$ 
obtained at $Q^2$ = 5 $({\rm GeV}/c)^2$ \cite{semiinc4}. 

In conclusion, the COMPASS collaboration has performed  new measurements of 
the longitudinal spin asymmetry of the proton, covering a large range of $x$
($0.004 < x < 0.7$) in the DIS region, $Q^2 > 1 ({\rm GeV}/c)^2$.
 The new data improve the statistical precision
in the low $x$ region by a factor of 2--3 and show no evidence  either for
an increase or a decrease 
 of the spin structure function $g_1^p$ in this region. In combination with the
previously published results on the deuteron, the new data improve the
evaluation of the non-singlet spin structure function $g_1^{NS}$ and provide
a test of the Bjorken sum rule, which is 
satisfied within one standard deviation of the statistical uncertainty.

\section*{Acknowledgements} 

We gratefully acknowledge the support of the CERN management and staff and 
the skill and effort of the technicians of our collaborating institutes. 
Special thanks go to V.~Anosov and V.~Pesaro for their technical support 
during the installation and the running of this experiment. 
This work was made possible thanks to the financial support of our funding agencies.

\section*{Appendix}
\begin{table}[here]
\centering
\begin{tabular}{|c|c|c|}
\hline  \hline
$x$ range    &Corr. to $A_1^p$ & Corr. to $A_1^d$ \\
\hline
0.004 - 0.005 & 0.000       &  0.000     \\
0.005 - 0.006 &  0.000       & 0.000      \\
0.006 - 0.008 &   0.000      & 0.001      \\
0.008 - 0.010 &  0.000       & 0.001       \\
0.010 - 0.020 &    0.000      &0.001       \\
0.020 - 0.030 &   0.000       &0.001        \\
0.030 - 0.040 &  -0.000       &0.002        \\
0.040 - 0.060 &  -0.001       & 0.002         \\
0.060 - 0.100 &  -0.001       & 0.003        \\
0.100 - 0.150 &  -0.003       &  0.004       \\
0.150 - 0.200 &   -0.004       & 0.006       \\
0.200 - 0.250 &   -0.005      &  0.008       \\
0.250 - 0.350 &  -0.006       &  0.010        \\
0.350 - 0.500 &  -0.009       &  0.013       \\
0.500 - 0.700 & -0.014       & 0.017       \\
\hline  \hline
\end{tabular}
\caption{\small Corrections to the COMPASS spin asymmetries $A_1^p$ and $A_1^d$ due to the
the $^{14}$N polarisation and to the admixture of $^7$Li and $^1$H into the $^6$LiD 
target material. 
In both cases the correction must be subtracted from the measured asymmetries.
The corrections to $A_1^p$ are already applied to the values quoted in the present letter.}
\label{tab:corr_A1p_A1d}
\end{table}

\end{document}